\documentclass[article,twocolumn,prl,showpacs]{revtex4-1}

\usepackage{graphicx}
\usepackage{amssymb}          
\usepackage[latin1]{inputenc} 
\usepackage{amsmath}          
\usepackage{color,ulem,todonotes}

\newcommand{\beq}{\begin{equation}}
\newcommand{\eeq}{\end{equation}}
\newcommand{\ket} [1] {|#1\rangle}
\newcommand{\bra} [1] {\langle#1|}
\newcommand{\braket} [2] {\langle#1|#2\rangle}

\newcommand{\I}{\Im{\rm m}}
\newcommand{\R}{\Re{\rm e}}
 
\begin{document}
\title{Strong measurements give a better direct measurement of the quantum wave function}
\author{Giuseppe Vallone}
\affiliation{Dipartimento di Ingegneria dell'Informazione, Universit\`a degli Studi di Padova, Padova, Italy.}
\author{Daniele Dequal}
\affiliation{Dipartimento di Ingegneria dell'Informazione, Universit\`a degli Studi di Padova, Padova, Italy.}

 \begin{abstract}
Weak measurements have thus far been considered instrumental in the so-called direct measurement
 of the quantum wavefunction
[Nature (London) 474, 188 (2011)].
Here we show that direct measurement of the wavefunction can be obtained by using measurements of arbitrary strength. In particular, 
in the case of strong measurements, i.e. those in which the coupling between the system and the measuring apparatus is maximum, 
we compared the precision and the accuracy of the two methods, 
by showing that strong measurements outperform weak measurements in both
for arbitrary quantum states in most cases.
We also give the exact expression of the difference between the reconstructed and original wavefunctions
obtained by the weak measurement approach: this
will allow to define the range of applicability of such method. 
\end{abstract} 
 \maketitle

{\it Introduction - }
In Quantum Mechanics the wavefunction is the fundamental representation of any quantum system,  and it offers the key tool for predicting the measurement outcomes of a physical apparatus. Its determination is therefore of crucial importance in many applications. In order to reconstruct the complete quantum wavefunction of a system, an indirect method, know as quantum state tomography (QST), has been developed~\cite{jame01pra}. QST is based on the measurement of complementary variables of several copies of the same quantum system, followed on an estimation of the wavefunction that better reproduce the results obtained. This method, originally proposed for a two level system, has been extended to a generic number of discrete quantum states~\cite{thew02pra} as well as to continuous variable state~\cite{lvov09rmp}. 
Recently Lundeen {\it et al.} \cite{lund11nat} proposed an alternative operational definition of the wavefunction based  on the weak measurement \cite{ahar88prl,ahar90pra,dres14rmp}.
After the first demonstration, in which the transverse wavefunction of a photon has been measured, this method has been applied for the measurement of the photon polarization~\cite{salv13nap}, its angular momentum~\cite{mali14nco} 
and its trajectory~\cite{kocs11sci}. The method has been subsequently generalized to mixed states~\cite{lund12prl} 
to continuous variable systems~\cite{fisc12pra} and compared to standard quantum state tomography in \cite{macc14pra,das14pra}.

By such method, that we call Direct-Weak-Tomography (DWT),  a ``direct measurement'' of the quantum wavefunction is obtained:
the term ``direct measurement'' refers to the property that a value proportional to the wavefunction 
appears straight from the measured probabilities
 without further complicated calculations or fitting on the measurement outcomes~\cite{foot1}.
As originally proposed~\cite{lund11nat}, the 
method is based on the weak-measurement obtained 
by a ``weak'' interaction between the ``pointer'' (i.e. the measurement apparatus) and system. 
Weak measurements occur when the coupling between the pointer and the system is much less than the pointer width. 
As reported in the literature,  {\it ``the crux of [the] method is that the first measurement is performed in a gentle way through weak measurement,
  so as not to invalidate the second''\cite{lund11nat}} or 
 {\it ``Directly measuring [...] relies on the technique of weak measurement: 
 extracting so little information from a single measurement that the state does not collapse''}\cite{salv13nap}.

The interest about DWT is that
the scheme in some cases may have experimental advantages over QST, in terms of simplicity, versatility, and directness~\cite{lund12prl}: 
it only requires a weak coupling of the system with an external pointer, a postselection of the final state of the system 
and a simple projective measurement of two complementary observables of the pointer, a two-level system. 
QST, in contrast, requires measuring a complete set of noncommuting observables of the system, 
which can be a very demanding requirement in systems with a large number of degrees of freedom. For instance 
the determination of the transverse spatial wavefunction of a single photon was first realized by DWT~\cite{lund11nat},
as well as the measurement of a one-million-dimensional photonic state~\cite{shi15qph}.

\begin{figure}[tbp]
\begin{center}
\includegraphics[width=7cm]{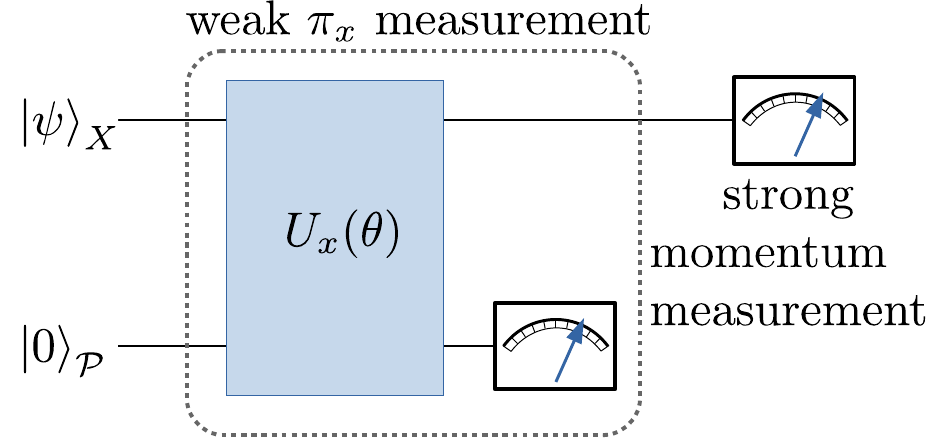}
\caption{Scheme of the original DWT method used to measure the wavefunction.}
\label{fig:scheme}
\end{center}
\end{figure}

Here we show that
the quantum wavefunction can be obtained by the same scheme used in DWT, but using only strong 
measurements:
with this terms we here refer
to measurements characterized by a strong coupling between the system and the pointer.
As explained below, a strong measurement does not always coincide with a projective measurement on the system.
 
 We thus demonstrate
that the weak measurement is not necessary for the direct measurement of the wavefunction. 
We then compare DWT with our method, showing that the use of strong measurements in most cases gives a better estimation of the quantum wavefunction, outperforming DWT when both accuracy and precision are considered.
Our analysis also allows to evaluate how  the wavefunction estimated by DWT is related to
the correct wavefunction, see eq. \eqref{psiW}. We also solved an unresolved question related to DWT: 
how ``weak'' the interaction should be such that DWT gives a correct estimation of the wavefunction. 
In particular, we will derive a sufficient criterium for the applicability of DWT based on the measured probabilities, see eq. \eqref{psiWcond}. 

{\it Review of Direct-Weak-Tomography - }
Let's consider a $d$ dimensional Hilbert space with basis $\{\ket{x}\}$ with $x=1,\dots, d$.
The states $\ket{x}$ are equivalent
to position eigenstates of a discretized segment.
A generic pure state in this basis can be written as
\beq
{\ket{\psi}}_X=\sum^d_{x=1}\psi_x\ket x \,.
\eeq
The scheme used in DWT is shown in figure \ref{fig:scheme}:
first, the following initial state $\ket{\Psi_{\rm in}}={\ket{\psi}}_X\otimes{\ket{0}}_{\mathcal P}$ is prepared,
with ${\ket{0}}_{\mathcal P}$ the {\it pointer} state. The pointer belongs to a bidimensional qubit space
spanned by the states 
$\{{\ket{0}}_{\mathcal P},{\ket{1}}_{\mathcal P}\}$~\cite{foot2}.
The system is then evolved according to the following unitary operator:
\beq
\label{interaction}
\begin{aligned}
U_x(\theta)&=e^{-i\theta\hat\pi_x\otimes \hat\sigma_y}\approx \openone-i\theta \,\hat\pi_x\otimes \hat\sigma_y\,,
\end{aligned}
\eeq
where $\theta$ is an arbitrary angle and $\hat\pi_x=\ket{x}\bra{x}$.
The approximation of the r.h.s. of eq. \eqref{interaction} is obtained for small $\theta$.
The previous evolution corresponds to a pointer rotation conditioned to $\ket{\psi}_X$ being in the state $\ket{x}$.
A projective measurement on the pointer, weakly coupled to the photon position and followed by a projective measurement of the photon momentum allows to directly determine the wavefunction.
Indeed, by post-selecting
only the outcomes corresponding to the zero transverse momentum state $\ket{p_0}=\frac{1}{\sqrt{d}}\sum_{x}\ket{x}$, 
the (unnormalized) pointer state becomes
${\ket{\varphi}}_{\mathcal P}\approx\frac{1}{\sqrt d}[\widetilde\psi{\ket{0}}_{\mathcal P}+\theta \psi_x{\ket 1}_{\mathcal P}]$
with $\widetilde \psi=\sum^d_{x=1}\psi_x$. 
The choice of $\ket{p_0}$ is arbitrary, and a different value of the transverse momentum might be needed for particular states, as explained below.
Since a global phase is not observable, it is possible to arbitrarily choose  the phase of $\widetilde \psi$:
we set the latter phase such that $\widetilde \psi$ is real valued and positive.
In the first order in $\theta$, 
the wavefunction can be derived directly as \cite{lund11nat}:
\beq
\label{psiW_simple}
\begin{aligned}
\psi_{W,x}
=\frac{d}{2\theta\,\widetilde\psi } \left[(P^{(x)}_+-P^{(x)}_-)+i(P^{(x)}_L-P^{(x)}_R)\right]\,,
\end{aligned}
\eeq
where $P^{(x)}_j$ represent
the probabilities of measuring the pointer state
into the diagonal  basis ${\ket{\pm}}_{\mathcal P}=\frac{1}{\sqrt2}(\ket 0\pm\ket 1)$,
or the circular basis
${\ket{L}}_{\mathcal P}=\frac{1}{\sqrt2}(\ket 0+i\ket 1)$ and ${\ket{R}}_{\mathcal P}=\frac{1}{\sqrt2}(\ket 0-i\ket 1)$.
We note that, since the (real positive) proportionality constant $\frac{d}{2\theta\widetilde\psi }$ is $x$ independent, 
it can be obtained at the end of the procedure by normalizing  the
 wavefunction. The different probabilities
 can be also expressed in the framework of POVM, as
 detailed is SI.
From now on, we indicate with $\psi_{W,x}$ the (approximate) wavefunction obtained with the DWT method.
We also define 
$\widetilde\psi_W\equiv\sum_x\psi_{W,x}=\frac{d}{2\theta\,\widetilde\psi }\sum_x(P^{(x)}_+-P^{(x)}_-)$
and we fix the global phase of $\psi_{W,x}$ by \eqref{psiW_simple}.

Relation \eqref{psiW_simple} was generalized to mixed states in \cite{lund12prl}. 
By repeating the measurements and changing the $x$ parameter in the evolution $U_x(\theta)$, the full wavefunction can be reconstructed.
We now show that a relation similar to \eqref{psiW_simple} can be obtained by strong or arbitrary strength measurements.

{\it Arbitrary strength measurement - }
Measurement with arbitrary strength is obtained by choosing arbitrary value of $\theta$ within $0<\theta\leq\pi/2$.
We start our analysis with strong measurements, corresponding to $\theta=\pi/2$. In this case the unitary operator \eqref{interaction} becomes $ U_x(\pi/2) = \openone- \ket{x}\bra{x} \otimes (\openone_\pi+i\sigma_y)$. 
After the interaction, the initial state  $\ket{\Psi_{\rm in}}$
is measured on the state 
$\ket{p_0}\otimes\ket{\phi_f}$,
where $\ket{\phi_f}$ is the
final polarization state.
 The amplitude for that transition is
just $\mathcal A =\braket{p_0}{\psi}_X 
\braket{\phi_f}{0}_\mathcal P  -\psi_x\sqrt{2/d} \braket{\phi_f}{-}_\mathcal P$.
This amplitude involves both the real and
imaginary parts of $\psi_x$, so its magnitude squared does too: by
choosing different values of $\ket{\psi_f}$, 
it is possible to determine the real and imaginary parts of $\psi_x$.
In particular by choosing the final state $\ket{\phi_f}$ as ${\ket1}_{\mathcal P}$, 
${\ket+}_{\mathcal P}$, ${\ket-}_{\mathcal P}$, ${\ket L}_{\mathcal P}$ and ${\ket R}_{\mathcal P}$ states, the wavefunction can be obtained as:
\beq
\label{strong_measure}
\begin{aligned}
\psi_x&=\frac{d}{2\widetilde\psi} \left[(P^{(x)}_+-P^{(x)}_-+2P^{(x)}_1)+i(P^{(x)}_L-P^{(x)}_R)\right].
\end{aligned}
\eeq
To obtain the above relation we fixed again $\widetilde \psi=|\widetilde \psi|$.
It is very important to stress that, differently from the DWT method,  the above result is exact, without any approximation.
We denote the previous relations as Direct-Strong-Tomography (DST) method.
The difference with respect to the DWT is the need of measuring the pointer state also in the state $\ket{1}_{\mathcal P}$. This extra requirement is compensated by the fact that the result is not approximated and the accuracy and precision of
the method overcomes the DWT, as we will show in the following. 
We underline that the measurement in the $\ket{1}_{\mathcal P}$ state, and only in this state, corresponds to a projective measurement of the photon position, as the outcome of the measurement is proportional to $|\psi_x|^2$ (see S.I.). On the contrary, a projection of the pointer in the $\{\ket+, \ket-\}$ or $\{\ket L, \ket R\}$ bases 
acts as a partial quantum erasure on the which-position information:
therefore a subsequent momentum
postselection allows to extract information about the real and imaginary part of $\psi_x$.
As detailed in SI, for arbitrary $\theta$, the wavefunction
can be obtained as 
$\Re e (\psi_x)\propto P^{(x)}_+-P^{(x)}_-+2\tan(\frac\theta2)P^{(x)}_1$ and
$\I (\psi_x)\propto P^{(x)}_L-P^{(x)}_R$.

{\it Accuracy of DWT - }
In the case of DWT, the obtained wavefuction $\psi_{W,x}$ is an approximation of the correct wavefunction $\psi_x$.
We now evaluate the accuracy of the DWT, namely the errors arising
 by using eq. \eqref{psiW_simple} in place of the exact values of \eqref{strong_measure}. 
As done in \cite{macc14pra}, we define the accuracy in terms of the trace distance $\mathcal D$ between the correct 
wavefunction $\psi_x$ and the weak value approximation $\psi_{W,x}$~\cite{foot3},
that for pure states 
reduces to $\mathcal  D= \sqrt{1-|\braket{\psi}{\psi_W}|^2}$.
We first give the analytical expression of $\mathcal  D$ in terms of the original wave function and
then show how $\mathcal D$ can be upper bounded by using the measurement outcomes.

As shown in SI, the relation between the exact wavefunction $\psi_x$ and the
weak-value estimate $\psi_{W,x}$ given in \eqref{psiW_simple} can be expressed by the following relation:
\beq
\label{psiW}
\psi_{W,x}=\psi_x\frac{\widetilde\psi-\epsilon_\theta\psi^*_x}{\mathcal N}\,,
\eeq
with 
$\epsilon_\theta\equiv2\sin^2(\frac\theta2)$,
$\mathcal N\equiv{\sqrt{|\widetilde\psi-\epsilon_\theta\langle \psi_x\rangle|^2+\epsilon_\theta^2\sigma^2_\psi}}$, 
and
$\sigma^2_{\psi}\equiv\langle |\psi_x|^2\rangle-|\langle \psi_x\rangle|^2$.
In the previous equation
$\sigma^2_{\psi}$ is the ``variance" of the wavefunction
where the average is defined with respect the probability density $p_x=|\psi_x|^2$, namely
$\langle|\psi_x|^2\rangle=\sum_x|\psi_x|^4$ and $\langle \psi_x\rangle=\sum_x \psi_x|\psi_x|^2$.
By inserting \eqref{psiW} into the trace distance $\mathcal D$ we obtain:
\beq
\label{D}
\mathcal D=\frac{\epsilon_\theta\sigma_\psi}{\mathcal N}\,,
\eeq
expressing $\mathcal D$ in terms of the original wavefunction $\psi_x$ and the interaction parameter $\theta$.
The previous expression indicates when the weak-measurement method can be efficiently used: indeed, when
\beq
\label{condition}
\mathcal D\ll 1\,,
\eeq
the approximate wavefunction $\psi_{W,x}$ correctly estimates the wavefunction $\psi_x$.
Since eq. \eqref{D} can be inverted into 
$\epsilon_\theta\sigma_\psi=\frac{\mathcal D}{\sqrt{1-D^2}}|\widetilde\psi-\epsilon_\theta\langle \psi_x\rangle|$,
for small $\mathcal D$ 
condition \eqref{condition} is equivalent to $\frac{\epsilon_\theta\sigma_\psi}{|\widetilde\psi-\epsilon_\theta\langle \psi_x\rangle|}\ll 1$
(see  SI for the detailed calculation).

\begin{figure}[tbp]
\begin{center}
\includegraphics[width=8cm]{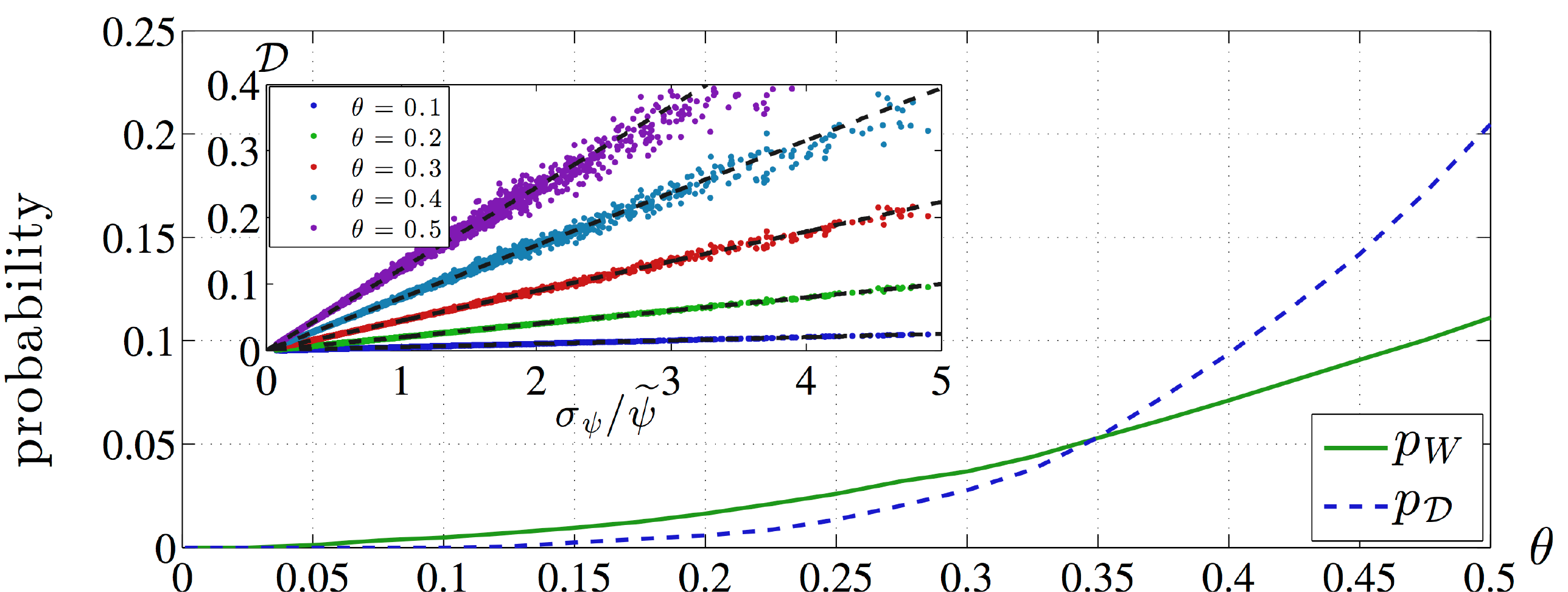}
\caption{Accuracy of the DWT: we show the
probability $p_W$ of having $\widetilde \psi_W<0$ and the probability $p_{\mathcal D}$ of having an error $\mathcal D$ larger that 0.1. 
The inset shows trace distance $\mathcal D$ in function of $\sigma_\psi/\widetilde\psi$ for different value of $\theta$. 
We randomly choose $10^6$ wavefunctions in a $d=10$ dimensional Hilbert space.
Dashed lines in the inset represent the curves $\mathcal D= \epsilon_\theta\frac{\sigma_\psi}{\widetilde\psi}$.
}
\label{fig:systematic}
\end{center}
\end{figure}

Condition \eqref{condition}, however,  cannot be used  if the exact wavefunction $\psi_x$ is unknown. 
For this reason, we now present 
a sufficient condition for the application of DWT method that is expressed in term of the measured probabilities. 
As shown in SI, when the follow inequality is satisfied
\beq
\label{psiWcond}
 \sum_x(P^{(x)}_+-P^{(x)}_-)
\geq 0\,,
\eeq
the systematic error is bounded by $\mathcal D\leq\theta/2$ (for small $\theta$).
We note that eq. \eqref{psiWcond}  is equivalent to
 $\widetilde\psi_W\geq0$ when  
the global phase of $\psi_{W,x}$ is fixed by eq. \eqref{psiW_simple}.

If condition \eqref{psiWcond} is not satisfied the DWT method is not guaranteed to work and a lower $\theta$ should be choosen to achieve condition 
\eqref{psiWcond}. 
Since $\widetilde\psi_W$ can be expressed in term of
the original wavefunction as $\widetilde\psi_W=\frac{\widetilde\psi^2-\epsilon_\theta}{\mathcal N}\,,$
for any wavefunction with $\widetilde\psi\neq0$ it is possible to lower $\theta$ such that condition \eqref{psiWcond}
is satisfied.
The wavefunctions with $\widetilde\psi=0$ corresponds to the set of ``pathological'' 
wavefunctions for which the DWT and the DST methods
can never be applied. 
Indeed, if $\widetilde \psi=0$ the systematic error \eqref{D} can be easily evaluated to be
$\mathcal D=\sigma_\psi/\sqrt{\langle|\psi_x|^2\rangle}$ that is independent of $\theta$: by changing the interaction parameter
the error cannot be lowered for such wavefunctions~\cite{foot4}.
Also for DST, the proportionality constant $\frac{d}{2\widetilde\psi\sin\theta}$ in \eqref{strong_measure} diverges if $\widetilde \psi=0$. 
In such case, a different momentum state for post-selection different from $\ket{p_0}$ must be used.

To better evaluate the accuracy  of the DWT we have randomly 
chosen $10^6$ wavefunctions in a $d=10$ dimensional Hilbert space according to the Haar measure.
We calculated for different values of $\theta$ the probability $p_W$ to violate the sufficient condition, namely 
$p_W={\rm Prob}(\widetilde\psi_W<0)$. We also calculated the probability $p_{\mathcal D}$ 
of having an error $\mathcal D$, evaluated by \eqref{D}, larger that 0.1. 
In Figure \ref{fig:systematic} we show the probabilities $p_W$ and $p_{\mathcal D}$ in function of $\theta$.
In the inset we also show the systematic error $\mathcal D$ in function of $\sigma_\psi/\widetilde\psi$ for different values of $\theta$.
Since the distribution of  $\mathcal N$ is peaked around $\widetilde\psi$ for $\theta\leq 0.5$,
it is possible to approximate $\mathcal D\approx \epsilon_\theta\frac{\sigma_\psi}{\widetilde\psi}$:
indeed, dashed lines in the inset of Fig. \ref{fig:systematic} represent the curves $\mathcal D= \epsilon_\theta\frac{\sigma_\psi}{\widetilde\psi}$.
The figure shows that for low values of $\theta$, the DWT method fails with low probability and
the systematic error is limited.
Indeed, if we choose $\theta\leq0.2$ for the $d=10$ case, we have $p_W\leq1.75\%$ and $p_\mathcal D\leq0.57\%$.
Then, as expected, low values of the interaction parameter $\theta$ are suitable for the correct application of the DWT method.
However, as we will show in the following, such low $\theta$ values lead to a larger statistical error (i.e. lower precision)
compared to the strong measurement method.

{\it Precision of the DWT -}
\begin{figure}[tbp]
\begin{center}
\includegraphics[width=8cm]{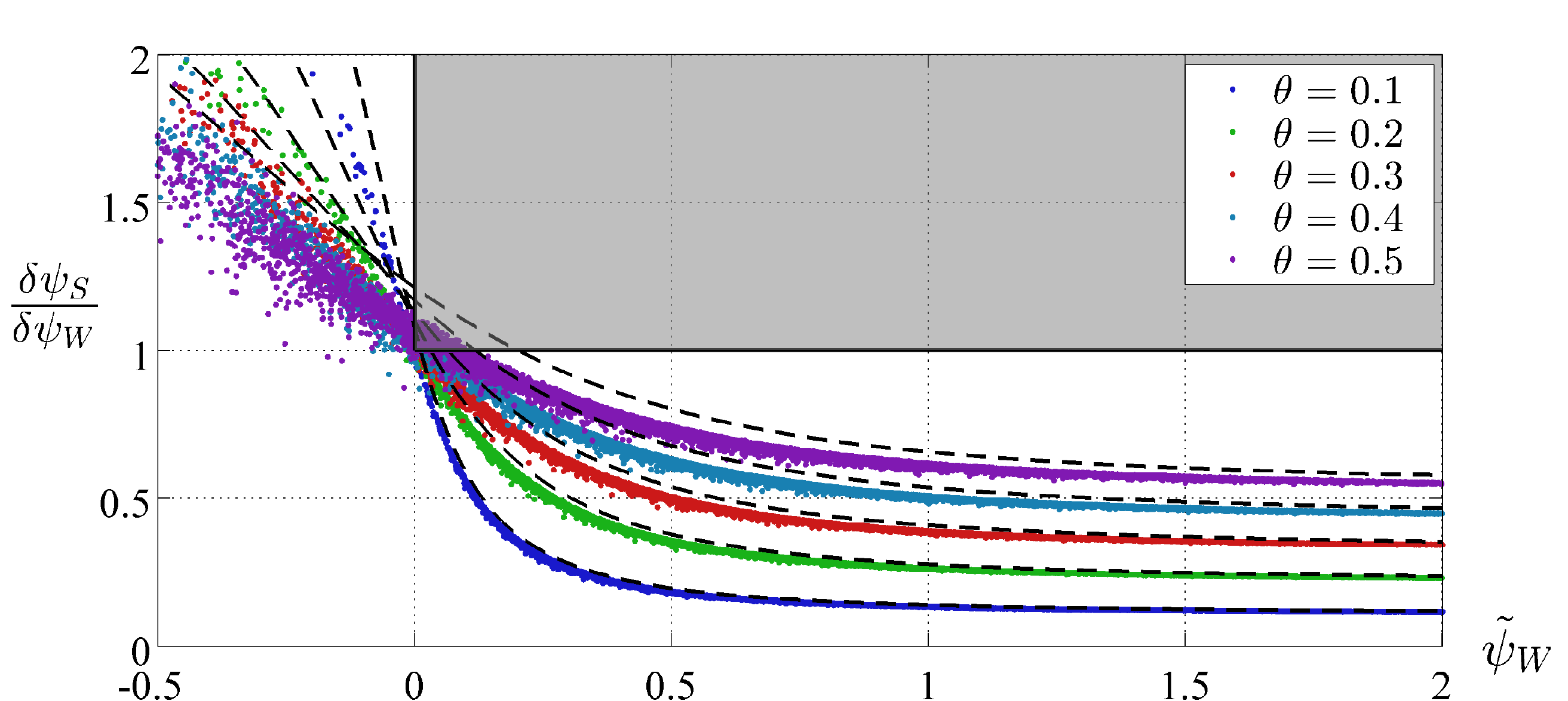}
\caption{Ratio of statistical errors $\frac{\delta\psi_S}{\delta \psi_W}$ in function of $\widetilde\psi_W$.
Shaded area represent the points in which the DWT is convenient with respect to the DST method,
corresponding to $\widetilde\psi_W\geq0$ and $\delta\psi_W\leq\delta\psi_S$.
}
\label{fig:ratio_psi}
\end{center}
\end{figure}
\begin{figure}[t!]
\begin{center}
\includegraphics[width=8cm]{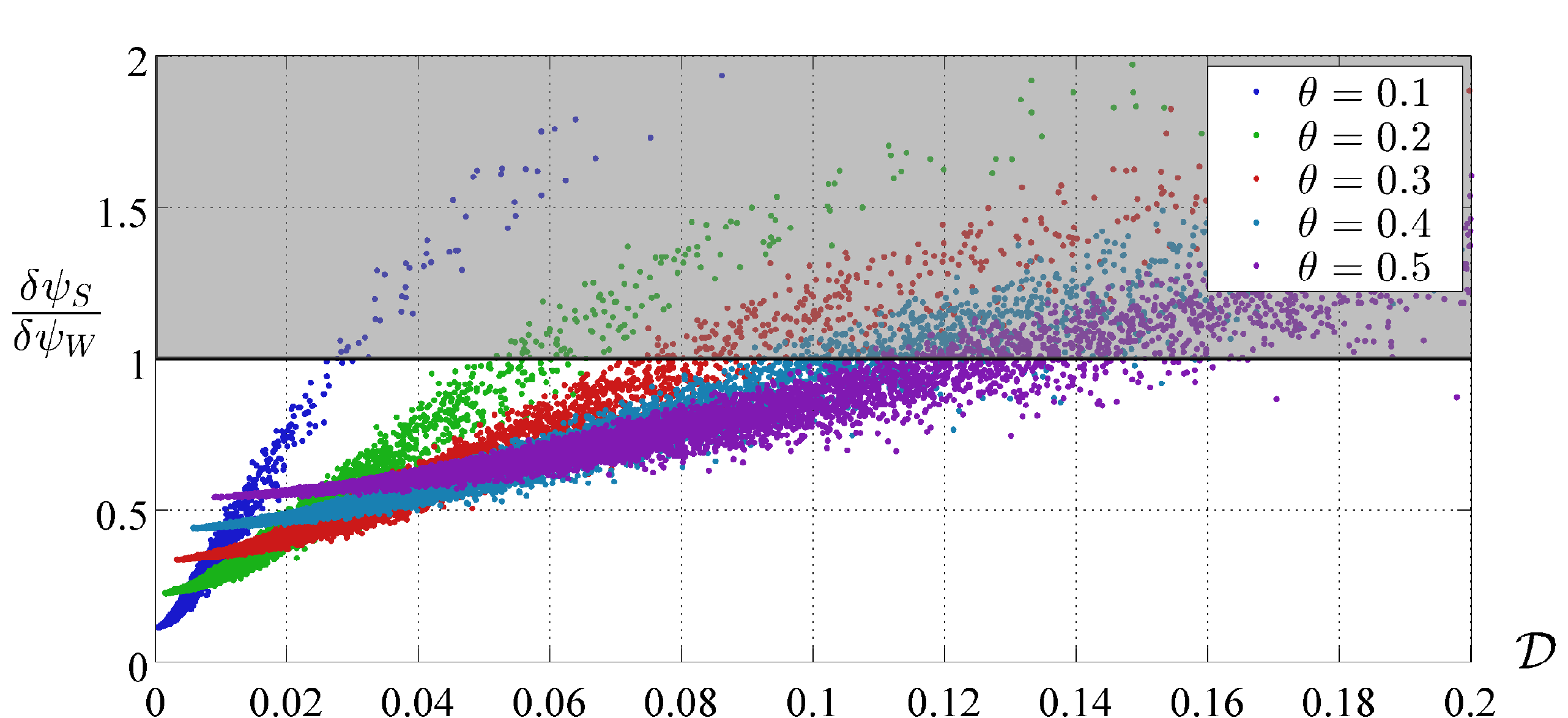}
\caption{Ratio of statistical errors $\frac{\delta\psi_S}{\delta \psi_W}$ in function of $\mathcal D$.
Shaded area represent the wavefuntions for which the statistical error of the DWT is lower than the  DST method.
}
\label{fig:ratio}
\end{center}
\end{figure}
An important performance parameter is the precision of the method, namely the statistical errors on the estimated wavefunction. 
In particular, it is important to evaluate the scaling of such errors with
the number of measurements. To this purpose, we 
evaluated the {\it mean square statistical error} $\delta \psi$ of the DWT and DST methods, 
obtained by summing the squares of the statistical error on the different $\psi_x$:
\beq
\delta\psi=\sqrt{\sum_x|\delta\psi_x|^2}\,.
\eeq
As shown in SI, the ratio between
the statistical errors $\delta\psi_S$ and $\delta\psi_W$, respectively corresponding to the strong and weak method,
 can be approximately bounded by:
\beq
\label{ratiodelta}
\frac{\delta\psi_S}{\delta \psi_W}
\lesssim
\sin\theta_0\sqrt{\frac{3}{2}}
\sqrt{\frac{(2d-5)\widetilde\psi^2+2\widetilde\psi + 8-2/d}
{(2d-1)\,\widetilde\psi^2 + 2\epsilon_\theta(1-\widetilde\psi-2\widetilde\psi^2)}}
\eeq
where $\theta_0$ is the interaction parameter used for the weak measurement.
The terms $\sin\theta_0$ in eq. \eqref{ratiodelta} shows
that low values of $\theta_0$ correspond to a lower precision (i.e. larger statistical errors) of the DWT with respect to the DST method.
In the statistical analysis, we compared the
two method by fixing the number of repetition $N$ of the experiment: in the DWT or DST method, $N/2$ or $N/3$ 
repetitions are used for each basis respectively.
This is the origin of the $\sqrt{3/2}$ factor in eq. \eqref{ratiodelta}.

For a complete demonstration of such feature we calculated the exact ratio $\frac{\delta\psi_S}{\delta \psi_W}$
for $10^6$ randomly chosen wavefunctions and compared it with the success parameter $\widetilde \psi_W$ and
 the systematic error $\mathcal D$. 
The results are shown in Fig. \ref{fig:ratio_psi} and \ref{fig:ratio}.
Figure  \ref{fig:ratio_psi} show that, when the sufficiency condition for applying the DWT is satisfied,
(i.e. $\widetilde\psi_W\geq0$), the statistical errors of the DWT are typically greater then the errors of the DST.
An approximate trent of the ratio ${\delta\psi_S}/{\delta \psi_W}$ can be obtained by noticing that, since $\mathcal N\approx\widetilde \psi$, 
we can approximate $\widetilde\psi_W\approx\widetilde\psi-\epsilon_\theta/\widetilde\psi$.
Dashed curves in Fig. \ref{fig:ratio_psi} represent the r.h.s. of eq. \eqref{ratiodelta}, with $\widetilde\psi$ replaced by
 $\frac12(\widetilde\psi_W+\sqrt{\widetilde\psi^2_W+4\epsilon_\theta})$ and well reproduce the behavior of the ratio $\delta\psi_S/\delta\psi_W$.
 
To further prove that the DST precision is typically greater than the DWT one,
we plot in Fig.  \ref{fig:ratio} the same ratio $\delta\psi_S/\delta\psi_W$ in function of the exact trace distance $\mathcal D$: 
for low systematic error $\mathcal D$, the statistical errors of the DWT are typically greater then the errors of DST.
Equivalently, statistical errors of the DWT are reduced only as the systematic errors increase.
Fig. \ref{fig:ratio} shows that the DST precision overcomes the DWT one in most of the cases in which the DWT is accurate.

\begin{figure}[t!]
\begin{center}
\includegraphics[width=8cm]{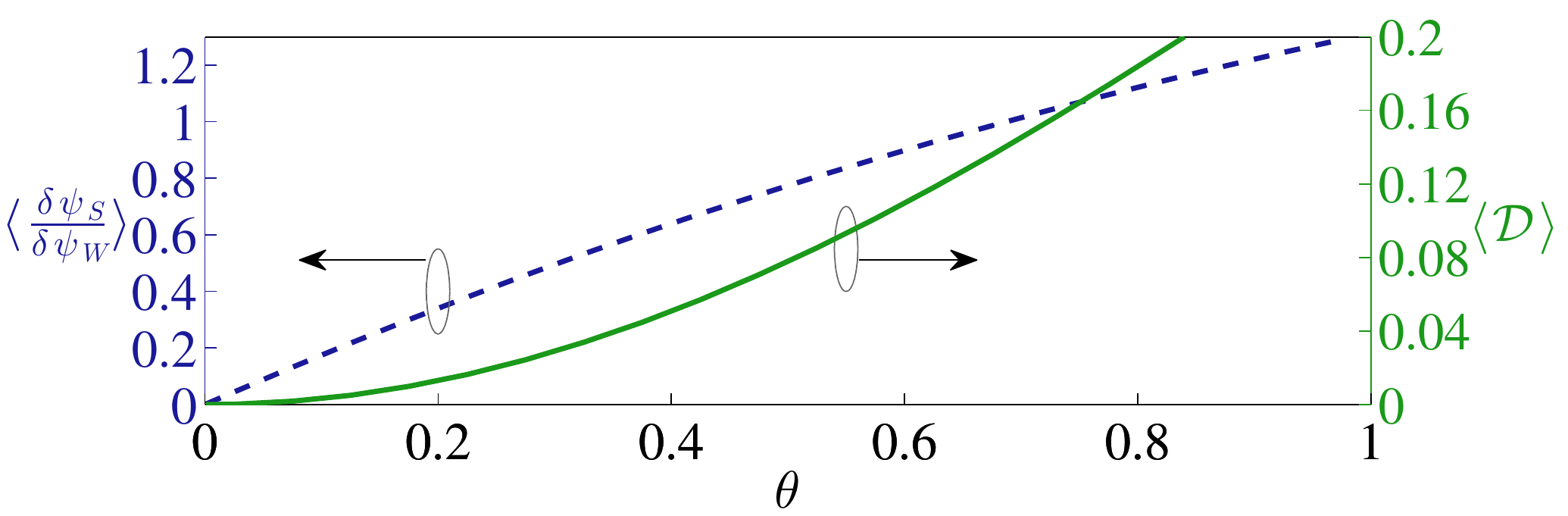}
\caption{
Mean values of 
$\frac{\delta\psi_S}{\delta\psi_W}$ and $\mathcal D$ averaged over $10^6$ random wavefunctions
in function of $\theta$.
}
\label{fig:prob}
\end{center}
\end{figure}
To better appreciate the above results, we  plot in Fig. \ref{fig:prob} the mean values of
$\frac{\delta\psi_S}{\delta\psi_W}$ and $\mathcal D$ averaged over $10^6$ random wavefunctions
in function of $\theta$.
The plot  in Fig. \ref{fig:prob} shows again that in order to lower the trace distance $\mathcal D$ it is necessary to
decrease $\theta$. However, decreasing $\theta$, the statistical error $\delta \psi_W$ becomes larger
than $\delta\psi_S$.

{\it Mixed states - }The DWT can be generalized to determine the density matrix $\rho$ of mixed states, as shown in \cite{lund12prl}.
To directly measure $\rho$ the same method described for pure state can be used, with the extra requirement
that the strong measurement on momentum should be performed in all the momentum states 
$\ket{p}=\frac{1}{\sqrt{d}}\sum_{x}e^{2\pi i \frac{p x}{d}}\ket{x}$, while the pointer is measured is the $\ket{\pm}_{\mathcal P}$, 
$\ket{R}_{\mathcal P}$, $\ket{L}_{\mathcal P}$ states (as done for the pure state $\ket{\psi}_X$). 
We indicate by $\rho^W$ the density matrix
that is reconstructed by the DWT 
and that approximates the correct matrix $\rho$. As shown in SI, it can be expressed as
\beq
\rho^{W}=\frac{1}{\cos\theta}\left[\rho+(\cos\theta-1)D\right]\,,
\eeq
with $D$  a diagonal matrix whose element are equal to the diagonal of $\rho$, namely $D_{x,y}=\delta_{x,y}\rho_{x,x}$.
By evaluating  the accuracy of the DWT in terms of the trace distance $\mathcal D$ between $\rho$ and $\rho^W$
we obtained
\beq
\mathcal D=
\frac{1-\cos\theta}{2\cos\theta}{\rm Tr}\left[\sqrt{(\rho-D)^2}\right]\,.
\eeq
Also in this case, the larger is $\theta$, the larger is $\mathcal D$ and the lower is the accuracy in the estimation of $\rho$ by the DWT.
Similarly to what we have shown for pure states, 
by performing an extra measurement of the pointer in the $\ket{1}_{\mathcal P}$ state, 
the exact expression of the density matrix can be obtained for any value of $\theta$ also in the case
of mixed states (see SI).

{\it Conclusions - } We have demonstrated that, in order to achieve a direct measurement of the wavefunction, weak measurements
are not necessary.
Indeed, we have shown that by using strong measurements, in which a large entanglement is achieved between the system and the pointer,
a better estimation of
the wavefunction, in terms of precision and accuracy, can be obtained
for random matrices in most cases.
Our method allowed us to derive a sufficient condition for the applicability of the Direct-Weak-Tomography.
We believe that our results give a deeper understanding of the meaning of the weak-value for 
the estimation of the wavefunction.

\begin{acknowledgements}
We thank {P. Villoresi of the University of Padova} and L. Maccone of the University of Pavia for useful discussions. 
Our work was supported by the 
Progetto di Ateneo PRAT 2013 (CPDA138592)
of the University of Padova.
G.V. also acknowledge the Strategic-Research-Project QUINTET of the Department of Information Engineering, University of Padova.
\end{acknowledgements}

\clearpage
\onecolumngrid
\appendix
\begin{center}
{\large\bf Supplementary information: \\
Strong measurements give a better direct measurement of the quantum wave function}
\end{center}

\maketitle

\setcounter{equation}{0}
\renewcommand{\theequation}{S\arabic{equation}}

\section{Derivation of the wavefunction by generic strength measurement }
\label{intermediate}
We here demonstrate the relation given in eq. (4) of the main text.
Let's consider a generic interaction parameter $\theta$ and
the input state
$\ket{\Psi_{\rm in}}={\ket{\psi}}_X\otimes{\ket{0}}_{\mathcal P}$. In this case the unitary operators $U_x(\theta)$ becomes
\beq
\begin{aligned}
U_x(\theta)&=\openone_x\otimes\openone_\pi-\ket{x}\bra{x}\otimes[(1-\cos\theta)\openone_\pi+i\sin\theta\sigma_y]\,,
\end{aligned}
\eeq
and the (unnormalized) pointer state after the momentum post-selection is given by
\beq
\begin{aligned}
{\ket{\varphi}}_{\mathcal P}
= \bra{p_0}U_x(\theta)\ket{\Psi_{\rm in}}
&=\frac{1}{\sqrt d}[\widetilde\psi\ket{H}+\psi_x \ket \chi]\,,
\end{aligned}
\eeq
were we have defined the (unnormalized) state $\ket{\chi}=(\cos\theta-1)\ket H+\sin\theta\ket V$ and
$\widetilde\psi=\sum_x\psi_x$.
As indicated in the main text, it is possible to choose the phase of the wave function such that
$\widetilde \psi=|\widetilde \psi|$.
By defining $\epsilon_\theta=2\sin^2\frac\theta2$,
the probabilities of measuring in the different pointer states are given by
\beq
\label{probabilities}
\begin{aligned}
P_0^{(x)}
=&\frac{1}d\left[\widetilde\psi^2-2\epsilon_\theta \widetilde\psi \R (\psi_x)+\epsilon^2_\theta|\psi_x|^2\right]
\approx\frac{\widetilde\psi^2}d
\\
P_1^{(x)}
=&\frac{1}d\sin^2\theta|\psi_x|^2\approx0
\\
P_+^{(x)}
=&\frac{1}{d}\left[\frac{\widetilde\psi^2}2-(\epsilon_\theta-\sin\theta)  \widetilde\psi\,\R (\psi_x)+(1-\sin\theta)\epsilon_\theta|\psi_x|^2
\right]
\approx\frac{\widetilde\psi}{d}\left[\frac{\widetilde\psi}2+\theta\,\R (\psi_x)\right]
\\
P_-^{(x)}
=&\frac{1}{d}\left[\frac{\widetilde\psi^2}2
-(\epsilon_\theta+\sin\theta)\widetilde\psi\,\R (\psi_x)
+(1+\sin\theta)\epsilon_\theta|\psi_x|^2
\right]
\approx\frac{\widetilde\psi}{d}\left[\frac{\widetilde\psi}2-\theta\,\R (\psi_x)\right]
\\
P_L^{(x)}
=&\frac{1}d\left[\frac{\widetilde\psi^2}2+\theta\widetilde\psi\,\I (\psi_x)+
\epsilon_\theta\left(|\psi_x|^2-\widetilde\psi\,\R (\psi_x)\right)\right]
\approx\frac{\widetilde\psi}{d}\left[\frac{\widetilde\psi}2
+\theta\I (\psi_x)\right]
\\
P_R^{(x)}
=&\frac{1}d\left[\frac{\widetilde\psi^2}2-\sin\theta\widetilde\psi\,\I (\psi_x)+
\epsilon_\theta\left(|\psi_x|^2-\widetilde\psi\,\R (\psi_x)\right)\right]
\approx\frac{\widetilde\psi}d\left[\frac{\widetilde\psi}2
-\theta\I (\psi_x)\right]
\end{aligned}
\eeq
For low $\theta$,
the approximate results of the
r.h.s. holds (at the first order in $\theta$). 

We note that, by defining $\alpha_j=\braket{e_j}{0}$ and $\beta_j=\braket{e_j}{1}$, the above probabilities can
be obtained by the following POVM:
\beq
\begin{aligned}
P_j&={\rm Tr}\left[E^\dag_jE_j{\ket{\psi}}_X\bra{\psi}\right]
\end{aligned}
\eeq
with
\beq
\begin{aligned}
E_j=\alpha_j\ket{p_0}\bra{p_0}-\gamma_j\ket{p_0}\bra{x}
\end{aligned}
\eeq
and $\gamma_j=\frac{1}{\sqrt{d}}\left[(1-\cos\theta)\alpha_j-\sin\theta\beta_j\right]$.

From the previous equations \eqref{probabilities}, 
using the exact results, it is possible to prove that:
\beq
\begin{aligned}
\Re e (\psi_x)
&=\frac{d}{2\widetilde\psi\sin\theta}[P_+^{(x)}-P_-^{(x)}+2\tan(\frac\theta2)P_1^{(x)}]\,,
\\
\I (\psi_x)&=\frac{d}{2\widetilde\psi\sin\theta}[P_L^{(x)}-P_R^{(x)}]\,.
\end{aligned}
\eeq
Strong measurements correspond to $\theta=\pi/2$.
By measuring the pointer in the $\ket{e_1}\equiv\ket1$, $\ket{e_+}\equiv\ket+$, $\ket{e_-}\equiv\ket-$, 
$\ket{e_L}\equiv\ket L$ and $\ket{e_R}\equiv\ket R$ basis, the wave function can be
thus derived. It is very important to stress that the result is exact, without any approximation.

If we consider the weak-value approximation, then the approximate values of $P_j^{(x)}$ can be used.
In this case
\beq
\begin{aligned}
\label{psiWx_SI}
\R(\psi_{W,x})&=\frac{d}{2\theta\widetilde\psi}(P^{(x)}_+-P^{(x)}_-)
\\
\I(\psi_{W,x})&=\frac{d}{2\theta\widetilde\psi}(P^{(x)}_L-P^{(x)}_R)
\end{aligned}
\eeq

\section{Relation between weak and strong value}
\label{psi-relation}
Let's now derive the the relation between the correct wave function $\psi_x$ and the
weak-value estimate $\psi_{W,x}$.
To evaluate the wave function it is necessary to estimate the parameters
$A_x=P_+^{(x)}-P_-^{(x)}+2\tan(\frac\theta2)P_V^{(x)}$ and $B_x=P_L^{(x)}-P_R^{(x)}$
such that the wavefunction is obtained by normalization: 
\beq
\psi_x=\frac{A_x+iB_x}{\mathcal M}
\eeq
with $\mathcal M=\sqrt{\sum_x(A^2_x+B^2_x)}=\frac{2\widetilde\psi\sin\theta}d$.
On the other hand, the weak value wave function $\psi_{W,x}$ is given by
\beq
\psi_{W,x}=\frac{A_{W,x}+iB_{W,x}}{\mathcal M_W}
\eeq
with the parameters given by
$A_{W,x}=P_+^{(x)}-P_-^{(x)}=\mathcal M [\R(\psi_x)-\frac{\epsilon_\theta}{\widetilde\psi}|\psi_x|^2]$,
$B_{W,x}=P_L^{(x)}-P_R^{(x)}=\mathcal M \I(\psi_x)$ and $\mathcal M_W=\sqrt{\sum_x(A_{W,x}^2+B_{W,x}^2)}$.
By comparing the two results we can write
\beq
\label{psiWx_SI}
\psi_{W,x}=\frac{1}{\mathcal N}\psi_x(\widetilde\psi-\epsilon_\theta\psi^*_x)
\eeq
with $\mathcal N$ determined by the normalization of $\psi_{W,x}$:
\beq
\begin{aligned}
\mathcal N
&=\sqrt{\sum_x|\psi_x|^2|\widetilde\psi-\epsilon_\theta\psi_x|^2}=\sqrt{\widetilde\psi^2
-2\epsilon_\theta\widetilde\psi\R\langle \psi_x\rangle+\epsilon^2_\theta
\langle|\psi_x|^2\rangle}
\\
&=\sqrt{|\widetilde\psi-\epsilon_\theta\langle \psi_x\rangle|^2+\epsilon_\theta^2\sigma^2_\psi}\,.
\end{aligned}
\eeq
The average is defined with respect the probability density defined by the wave function, $p_x=|\psi_x|^2$, namely
$\langle|\psi_x|^2\rangle=\sum_x|\psi_x|^4$ and $\langle \R(\psi_x)\rangle=\sum_x \R(\psi_x)|\psi_x|^2$.
We have thus demonstrated eq. (5) of the main text.

We now show that the trace distance between $\psi_{W,x}$ and $\psi_{x}$ can be bounded by knowing $\psi_{W,x}$.
The trace distance can be  exactly evaluated if we know the correct wave function $\psi_x$, by
\beq
\mathcal D=\frac{\epsilon_\theta\sigma_\psi}{\mathcal N}\,.
\eeq
The relation between $\mathcal D$ and $\epsilon_\theta\sigma_\psi$ can be inverted by squaring the previous equation,
namely $\mathcal D^2=\frac{(\epsilon_\theta\sigma_\psi)^2}{|\widetilde\psi-\epsilon_\theta\langle \psi_x\rangle|^2+(\epsilon_\theta\sigma_\psi)^2}$.
By resolving for $\epsilon_\theta\sigma_\psi$ we obtain $\epsilon_\theta\sigma_\psi=\frac{\mathcal D}{\sqrt{1-D^2}}|\widetilde\psi-\epsilon_\theta\langle \psi_x\rangle|$ that for low $\mathcal D$ can be approximated by
$\epsilon_\theta\sigma_\psi\approx{\mathcal D}|\widetilde\psi-\epsilon_\theta\langle \psi_x\rangle|$,
such that ${\mathcal D}\approx\frac{\epsilon_\theta\sigma_\psi}{|\widetilde\psi-\epsilon_\theta\langle \psi_x\rangle|}$.
Then,
for small $\mathcal D$, condition $\mathcal D\ll1$ is equivalent to $\frac{\epsilon_\theta\sigma_\psi}{|\widetilde\psi-\epsilon_\theta\langle \psi_x\rangle|}\ll 1$.

The parameter $\mathcal D$ can be bounded by knowing $\widetilde\psi_W\equiv\sum_x\psi_{W,x}$. 
We note that the global phase of $\psi_{W,x}$ is fixed
by \eqref{psiWx_SI}.
By using
\eqref{psiWx_SI}, we have $\widetilde\psi_W=(\widetilde\psi^2-\epsilon_\theta)/\mathcal N$. 
When $\widetilde\psi_W\geq0$ we can conclude that
$\widetilde\psi\geq\sqrt{\epsilon_\theta}$ allowing to bound the parameter $\mathcal D$. Indeed,
since $|\langle \psi_x\rangle|\leq 1$ and $\epsilon_\theta\leq 1$,
the condition $\widetilde\psi\geq\sqrt{\epsilon_\theta}$ implies $|\widetilde\psi|\geq\epsilon_\theta|\langle\psi_x\rangle|$ and
\beq
 |\widetilde\psi-\epsilon_\theta\langle \psi_x\rangle|^2\geq (\widetilde\psi-\epsilon_\theta|\langle \psi_x\rangle|)^2
 \geq( \widetilde\psi-\epsilon_\theta)^2
\geq \epsilon_\theta(1-\sqrt{\epsilon_\theta})^2\,.
\eeq
Finally, since $\sigma_\psi\leq1/\sqrt2$ we can conclude that
\beq
\label{boundSI}
\widetilde\psi_W\geq0
\qquad\Rightarrow
\qquad
\mathcal D=\frac{1}{\sqrt{\frac{|\widetilde\psi-\epsilon_\theta\langle \psi_x\rangle|^2}{\epsilon^2_\theta\sigma^2_\psi}+1}}
\leq\frac{1}{\sqrt{2\frac{(1-\sqrt{\epsilon_\theta})^2}{\epsilon_\theta}+1}}
\leq\frac{\sqrt{\epsilon_\theta}}{\sqrt{2-4\sqrt{\epsilon_\theta}+3\epsilon_\theta}}
\approx\theta/2
\eeq
where the last approximate result holds for low
$\theta$.
The previous relation proves that the condition $\widetilde\psi_W\geq0$ gives an upper bound on the systematic
error $\mathcal D$. 
By eq. \eqref{psiWx_SI} the sign of $\widetilde\psi_W$ is equal to the
sign of $\sum_x(P^{(x)}_+-P^{(x)}_-)$: then equation
\eqref{boundSI} proves eq.
(8) of the main text.

\section{Analysis of the precision of the DWT and DST methods}
It is useful to introduce the following {\it average error} $\delta \psi$, obtained by summing the squares of the statistical
error on the different $\psi_x$:
\beq
\delta\psi=\sqrt{\sum_x|\delta\psi_x|^2}
\eeq
with $|\delta\psi_x|^2=\delta\R(\psi_x)^2+\delta\I(\psi_x)^2$.

In general, for a wavefunction written as $\psi_x=\frac{A_x+iB_x}{\mathcal M}$ with $\mathcal M=\sqrt{\sum_x(A^2_x+B^2_x)}$ we have

\beq
\label{deltapsi}
\delta \psi=
\sqrt{\sum_x[(1-\frac{A_x^2}{\mathcal M^2})\delta A^2_x+(1-\frac{B_x^2}{\mathcal M^2})\delta B^2_x]}
\eeq

Let's now evaluate the above expression for the two methods, the DWT and the DST.

\subsection{DWT}
Let's now evaluate such average error $\delta\psi$ in the weak measurement case. 
Let's consider to repeat the experiment $N$ times.
For the weak value we need to measure in the $\{\ket{L},\ket{R}\}$\ and the 
$\{\ket{+},\ket{-}\}$ basis. Let's suppose that $N/2$ measurements are used for the first basis and $N/2$ per the remaining basis.
We indicate with tilde the estimated parameters obtained ofter $N$ measurements. 
The estimate for the polarization probabilities $P_j$ are
\beq
\widetilde P_j=\frac{n_j}{N/2}\xrightarrow{\quad N\rightarrow\infty\quad}P_j
\eeq
since $n_j\rightarrow P_jN/2$ in the large $N$ limit. From now on we indicate with a right arrow the asymptotic behavior
in the large $N$ limit.
The variance of $n_j$ is equal to $n_j$ due to Poissonian statistic.
Then
\beq
\delta\widetilde P_j=\frac{2\delta n_j}{N}=\frac{2\sqrt{n_j}}{N}\xrightarrow{\quad }\sqrt{\frac{2P_j}{N}}
\eeq
The probabilities are used to estimate the terms
\beq
\widetilde A_{W,x}=\widetilde P_+-\widetilde P_-\,,\quad
\widetilde B_{W,x}=\widetilde P_L-\widetilde P_R\,,\quad
\eeq
from which the wave function is obtained in the large $N$ limit as $\R(\psi_x)=\frac{\widetilde A_{W,x}}{\widetilde{\mathcal M}_W}$ and
$\I(\psi_x)= \frac{\widetilde B_{W,x}}{\widetilde {\mathcal M}_W}$ with the  factor $\widetilde{\mathcal M}_W$  determined by
normalization
$\widetilde{\mathcal M}_W=\sqrt{\sum_x(\widetilde A^2_{W,x}+\widetilde B^2_{W,x})}$.
In the large $N$ limit, the estimated $\widetilde{\mathcal M}_W$ approaches to 
$\widetilde{\mathcal M}_W\rightarrow \frac{2\sin\theta}{d}\mathcal N$.
The statistical error on the estimated $\widetilde A_x$ and $\widetilde B_x$ are given by
\beq
\label{deltaAW}
\begin{aligned}
\delta A_{W,x}&=\sqrt{\delta^2\widetilde P_++\delta^2\widetilde P_-}\xrightarrow
{\quad }\sqrt{\frac{2}{dN}}\sqrt{\widetilde\psi^2-2\epsilon_\theta\widetilde\psi\R(\psi_x)+2\epsilon_\theta|\psi_x|^2}
\\
\delta B_{W,x}&=\sqrt{\delta^2\widetilde P_L+\delta^2\widetilde P_R}\xrightarrow{\quad }
\delta A_{W,x}
\end{aligned}
\eeq

Since in the large $N$ limit we have $\frac{\widetilde A_{W,x}}{\widetilde {\mathcal M}_W}\rightarrow \R(\psi_{W,x})$, 
$\frac{\widetilde B_{W,x}}{\widetilde {\mathcal M}_W}\rightarrow \I(\psi_{W,x})$,
the mean square statistical error $\delta \psi_W$ is given by 
\beq
\label{deltapsiW}
\delta \psi_W=\frac1{\mathcal M_W}
\sqrt{\sum_x[(1-\R(\psi_{W,x})^2)\delta A^2_{W,x}+(1-\I(\psi_{W,x})^2)\delta B^2_{W,x}]}
\eeq
From \eqref{deltaAW} we have
\beq
\begin{aligned}
\sum_x(\delta A^2_{W,x}+\delta B^2_{W,x})
&=\frac{2}{dN} \left[ 2d\,\widetilde\psi^2 + 4\epsilon_\theta(1-\widetilde\psi^2)\right]
\\
\sum_x[\R(\psi_{W,x})^2\delta A^2_{W,x}+\I(\psi_{W,x})^2\delta B^2_{W,x}]
&=\frac{2}{dN} \left\{\sum_x|\psi_{W,x}|^2
(\widetilde\psi^2-2\epsilon_\theta\widetilde\psi\R(\psi_x)+2\epsilon_\theta|\psi_x|^2)\right\}
\end{aligned}
\eeq

By using the previous equation in \eqref{deltapsiW} we obtain, for the weak value case,
\beq
\begin{aligned}
\delta\psi_W\rightarrow
&\frac1{\sin\theta }\frac{1}{\mathcal N}
\sqrt{\frac{d}{2N}}\sqrt{(2d-1)\,\widetilde\psi^2 + 4\epsilon_\theta(1-\widetilde\psi^2)+2\epsilon_\theta
\sum_x|\psi_{W,x}|^2
(\widetilde\psi\R(\psi_x)-|\psi_x|^2)}
\\
&\geq\frac1{\sin\theta }\frac{1}{\mathcal N}
\sqrt{\frac{d}{2N}}\sqrt{(2d-1)\,\widetilde\psi^2 + 2\epsilon_\theta(1-\widetilde\psi-2\widetilde\psi^2)}
&\end{aligned}
\eeq
where we used $-\sum_x|\psi_{W,x}|^2|\psi_x|^2\geq -1$ and $\widetilde\psi\sum_x|\psi_{W,x}|^2\R(\psi_x)\geq-\widetilde\psi$.

\subsection{DST}
Let's now evaluate the average error $\delta\psi$ in the strong measurement case. 
Again we consider to repeat the experiment $N$ times.
We note that in this case we need to measure the ancillary qubit in three bases, namely $\{\ket{L},\ket{R}\}$,
$\{\ket{+},\ket{-}\}$ and $\{\ket{0},\ket{1}\}$ basis. 
Then, $N/3$ measurements are used for each basis, such that
\beq
\delta\widetilde P_j\xrightarrow{\quad }\sqrt{\frac{3P_j}{N}}\,.
\eeq
The $A_x$ and $B_x$ factor in eq. \eqref{deltapsi}
are given by:
\beq
A_x=\widetilde P_+-\widetilde P_-+2\tan(\frac\theta2)\widetilde P_V\,,
\quad
B_x=\widetilde P_L-\widetilde P_R\,,
\eeq
with errors
\beq
\label{deltaABstrong}
\begin{aligned}
\delta A_x&=\sqrt{\delta^2\widetilde P_++\delta^2\widetilde P_-+4\tan^2(\frac\theta2)\delta^2\widetilde P_V} 
\xrightarrow{\quad}\sqrt{\frac{3}{Nd}}\sqrt{\widetilde\psi^2-2\epsilon_\theta\widetilde \psi\Re e(\psi_x)  +2|\psi_x|^2\epsilon_\theta
(1+2\epsilon_\theta)}
\\
\delta B_x&=\sqrt{\delta^2\widetilde P_L+\delta^2\widetilde P_R}
\xrightarrow{\quad}\sqrt{\frac{3}{dN}}\sqrt{\widetilde\psi^2-2\epsilon_\theta\widetilde\psi\R(\psi_x)+2\epsilon_\theta|\psi_x|^2}\,.
\end{aligned}
\eeq
In the large $N$ limit we have $\frac{\widetilde A_{x}}{\widetilde {\mathcal M}}\rightarrow \R(\psi_{x})$, 
$\frac{\widetilde B_{x}}{\widetilde {\mathcal M}}\rightarrow \I(\psi_{x})$,
and the mean square statistical error $\delta \psi$ become
\beq
\label{deltapsi}
\delta \psi=\frac1{\mathcal M}
\sqrt{\sum_x[(1-\R(\psi_{x})^2)\delta A^2_{x}+(1-\I(\psi_{x})^2)\delta B^2_{x}]}
\eeq

By using \eqref{deltaABstrong} we have
\beq
\begin{aligned}
\sum_x(\delta A^2_x+\delta B^2_x)
&=\frac{3}{Nd} \left[ 2d\widetilde\psi^2 + 4\epsilon_\theta(1+\epsilon_\theta-\widetilde\psi^2)\right]
\\
\sum_x[\R(\psi_x)^2\delta A^2_x+\I(\psi_x)^2\delta B^2_x]
&=\frac{3}{Nd} \left[\widetilde\psi^2-2\epsilon_\theta\widetilde \psi\Re e\langle\psi_x\rangle+2\epsilon_\theta\langle|\psi_x|^2\rangle 
+4\epsilon^2_\theta \langle\Re e(\psi_x)^2\rangle\right]
\end{aligned}
\eeq
In the large $N$ limit we have $\mathcal M\rightarrow\frac{2\widetilde\psi\sin\theta}{d}$
such that
\beq
\begin{aligned}
\delta \psi
&=\frac{1}{\widetilde\psi\sin\theta}\sqrt{\frac{3d}{4N}}
\sqrt{(2d-1)\widetilde\psi^2 + 4\epsilon_\theta(1+\epsilon_\theta-\widetilde\psi^2)
+2\epsilon_\theta\widetilde\psi\Re e\langle\psi_x\rangle-2\epsilon_\theta\langle|\psi_x|^2\rangle 
-4\epsilon^2_\theta \langle\Re e(\psi_x)^2\rangle}
\end{aligned}
\eeq
For the strong measurement we have $\theta=\pi/2$ and $\epsilon_\theta=1$ such that
\beq
\begin{aligned}
\delta \psi_S
&=\frac{1}{\widetilde\psi}\sqrt{\frac{3d}{4N}}\sqrt{(2d-5)\widetilde\psi^2 + 8
- 2 \langle|\psi_x|^2\rangle
-4\langle  \Re e(\psi_x)^2\rangle+2\widetilde\psi\,\langle  \Re e(\psi_x)\rangle}
\\
&\leq\frac{1}{\widetilde\psi}\sqrt{\frac{3d}{4N}}\sqrt{(2d-5)\widetilde\psi^2+2\widetilde\psi + 8-2/d}
\end{aligned}
\eeq
since {$\langle\Re e(\psi_x)\rangle\leq1$, $\langle|\psi_x|^2\rangle\geq\frac{1}{d}$ and $\widetilde\psi\leq \sqrt{d}$}.

The ratio between the statistical errors can be bounded by 
\beq
\begin{aligned}
\frac{\delta\psi_S}{\delta \psi_W}
&\leq\sin\theta\sqrt{\frac{3}{2}}\frac{\mathcal N}{\widetilde\psi}
\sqrt{\frac{(2d-5)\widetilde\psi^2+2\widetilde\psi + 8-2/d}
{(2d-1)\,\widetilde\psi^2 + 2\epsilon_\theta(1-\widetilde\psi-2\widetilde\psi^2)}}
\end{aligned}
\eeq
that is the main result (eq. (9) of the main text) due to the fact that $\frac{\mathcal N}{\widetilde\psi}$ is well peaked around 1.
For large $\widetilde\psi$ we have $\mathcal N\sim\widetilde\psi\sim\widetilde\psi_W$ and the bound is simplified to
$\frac{\delta\psi_S}{\delta \psi_W}\sim\sin\theta\sqrt{\frac{3}{2}}\sqrt{\frac{2d-5}{2d-1}}$

\section{Mixed states with intermediate measure}
Let's consider the system initially prepared in the state
\begin{equation}
\label{mixed_initial_state}
\rho = \rho_X \otimes \ket{0}_{\mathcal P} \bra{0} \,,\qquad {\rm with}\quad \rho_X=\sum\limits_{x,y=1} ^{n} \hat{\rho}_{x,y} \ket{x} \bra{y}
\end{equation}
and $\ket{0}_{\mathcal P}$ the initial ``pointer'' state.
We would like to estimate the density matrix $\rho_X$.
After the interaction with $U_x(\theta)=e^{-i\theta\ket{x}\bra{x}\otimes \sigma_y}$, the state becomes
\begin{equation}
\begin{split}
\rho'_x  \equiv  U_x(\theta) \rho \, U^\dag_x(\theta) 
\end{split}
\end{equation}

The system is then measured into the momentum state $ \ket{p} = \frac{1}{\sqrt d} \sum\limits_{y=1} ^{d}  e^{\frac{2\pi i y p}{d}} \ket y $
such that the remaining ``pointer''  becomes:
\begin{equation}
\begin{split}
\rho ^{\mathcal P}_{x,p}
 \equiv \bra{ p} \rho'_x \ket{p}
= 
\left(
\begin{matrix}
\rho_{00}( x, p) & \rho_{01}(x, p)
\\
\rho_{10}(x, p) & \rho_{11}(x,p)
\end{matrix}
\right)
\end{split}
\end{equation}
with elements
$\rho_{00}( x, p) = \frac{1}{d} \left[ \sum_{x,y} \hat \rho_{x,y}  e^{\frac{2\pi i (y-x) p}{d}}  
- 2\sin^2\frac\theta2\sum_y \left(\hat \rho_{x,y} e^{\frac{2\pi i (y-x) p}{d}} +c.c.\right)
+ 4\sin^4\frac\theta2\hat \rho_{x,x} \right]$,
$\rho_{10}(x, p) = \frac{1}{d} \sin\theta\left[ \sum_y \hat \rho_{x,y} e^{\frac{2\pi i (y-x) p}{d}} 
-2\sin^2\frac\theta2\hat \rho_{x,x} \right]$,
$\rho_{01}( x, p) = \rho_{10}(x, p)^*$ and
$\rho_{11}(x,p) \frac{1}{d}  \sin^2\theta\hat \rho_{ x, x}$.

Now it possible to determine $\hat\rho_{x,y}$ in function of the ``pointer'' density matrix as follows:
\begin{gather}
\label{pointer_density}
\hat \rho_{x, y}
\propto
d\tan\frac\theta2 \delta_{ x, y}\rho_{11}( x, p) +\sum_{ p} e^{\frac{2\pi i ( x- y) p}{d}}\rho_{10}( x, p)
\\
\notag\downarrow{\rm\ strong\ measure\ }(\theta=\pi/2) 
\\
\label{rho_exact}
\hat \rho_{x, y}\propto
d \delta_{ x, y}\rho_{11}( x, p) +\sum_{ p} e^{\frac{2\pi i ( x- y) p}{d}}\rho_{10}( x, p) \,.
\end{gather}
The weak value estimate is obtained at the lowest order in $\theta$:
\beq
\label{rho_W}
\begin{aligned}
\hat \rho^W_{ x, y}&\propto
\left[\sum_{ p} e^{\frac{2\pi i ( x- y) p}{d}}\rho_{10}( x, p)\right]
\end{aligned}
\eeq
By the above equation it is possible to express the estimated density matrix $\rho^{W}$ in terms of the correct density $\rho$ as
\beq
\rho^{W}=\frac{1}{\cos\theta}\left[\rho+(\cos\theta-1)D\right]
\eeq
with $D$  a diagonal matrix whose element are equal to the diagonal of $\rho$, namely $D_{x,y}=\delta_{x,y}\rho_{x,x}$.

To determine the terms  $\rho_{10}$ in \eqref{rho_W} it is necessary to 
measure the pointer into four states $\ket{+}_{\mathcal P}$, $\ket{-}_{\mathcal P}$, $\ket{L}_{\mathcal P}$ and $\ket{R}_{\mathcal P}$.
Indeed, by defining $P_i^{(x,p)}\equiv\bra{i}\rho^{\prime\prime}_{x,p}\ket{i}$, we have the DWT relations:
\beq
\begin{aligned}
\rho_{10}( x, p) &=\frac12\left[(P^{( x, p)}_+-P^{( x, p)}_-)-i(P^{( x, p)}_L-P^{( x, p)}_R)\right]
\end{aligned}
\eeq

The DST allows to determine the exact density matrix $\rho$ by also measuring the pointer into the state $\ket{1}_{\mathcal P}$.
Indeed, to determine $\rho$ by eq. \eqref{rho_exact} it is necessary to evaluate also the 
the terms  $\rho_{11}$. It is easy to show that such terms can be evaluated by measuring the pointer into the states $\ket{1}_{\mathcal P}$:
\beq
 \rho_{11}( x, p)=P^{( x, p)}_{1} \equiv P^{( x)}_{1} 
\eeq

To summarize, for mixed states the procedure
is similar to the one performed with pure state.
The initial state \eqref{mixed_initial_state} is
evolved according to the interaction $U_x(\theta)$ between
the system and the pointer state. The system state is
strongly measured in a given momentum state $\ket{p}$
such that the pointer is left into a mixed two-level
system $\rho^{\mathcal P}_{x,p}$ given by eq. \eqref{pointer_density}.
By performing a standard tomography on the pointer,
namely by projecting it into 
$\ket{+}_{\mathcal P}$, $\ket{-}_{\mathcal P}$, $\ket{L}_{\mathcal P}$ and $\ket{R}_{\mathcal P}$
the pointer state $\rho^{\mathcal P}_{x,p}$ can
be obtained. Then, the exact initial density
matrix can be derived by eq. \eqref{pointer_density}.


\begin{thebibliography}{10}
\bibitem{foot1}
We used the
 term ``direct measurement'' to identify the
 method proposed \cite{lund11nat}. As illustrated in the SI, the method can be described in the more general framework of POVMs.
\bibitem{foot2}
In the case of  photon spatial wavefunction, the pointer can be represented by 
a different degrees of freedom of the photon, such as the polarization.
\bibitem{foot3}
We here recall that
the trace distance between two quantum states $\rho$ and $\rho'$ is defined as $\mathcal D\equiv\frac12\text{Tr}[|\rho-\rho'|]$.
\bibitem{foot4}
For example by using the DWT method on 
the following wavefunction $\psi_1=1/\sqrt{2}$, $\psi_2=-1/\sqrt{2}$ and $\psi_x=0$ for $x>2$ the
statistical error is maximal, $\mathcal D=1$, for any value of $\theta$.
\end{thebibliography}
\end{document}